\documentstyle[aasms4]{article}

\lefthead{Grebel \& Guhathakurta}
\righthead{Properties of new M31 companions}

\begin{document}

\title{Properties of Two New M31 dSph Companions from Keck Imaging
}

\author{Eva K.\ Grebel\altaffilmark{1,2} 
and Puragra Guhathakurta\altaffilmark{1,3,4}}

\affil{$^1$UCO/Lick Observatory, University of California at Santa Cruz,
Santa Cruz, California 95064, USA}
\authoremail{grebel@ucolick.org, raja@ucolick.org}
\affil{$^2$Hubble Fellow}
\affil{$^3$Visiting Astronomer, W.$\,$M.\ Keck Observatory, jointly operated
by the California Institute of Technology and the University of California}
\affil{$^4$Alfred P.\ Sloan Research Fellow}

\begin{abstract} 
We have obtained Keck Low Resolution Imaging Spectrograph images in $V$ and
$I$ of the newly discovered Local Group dwarf galaxies Pegasus\,dSph and
Cassiopeia\,dSph and their surrounding field.  The first stellar luminosity
functions and color-magnitude diagrams are presented for stars with
$V\lesssim25$ and $I\lesssim24$.  The distances to the new dwarfs are
estimated from the apparent $I$~mag of the tip of the red giant branch to be:
$D_{\rm MW}(\rm Peg\,dSph)=830\pm80$~kpc and $D_{\rm MW}(\rm
Cas\,dSph)=760\pm70$~kpc, consistent with their belonging to the extended M31
satellite system.  Both galaxies are dwarf spheroidals (dSphs) with red giant
branch morphologies indicating predominantly old stellar populations and
estimated mean metallicities, $\rm\langle[Fe/H]\rangle$, of $-1.3\pm0.3$ for
Peg\,dSph and $-1.4\pm0.3$ for Cas\,dSph.  Their central surface
brightness--$\rm\langle[Fe/H]\rangle$--absolute magnitude relationship
follows that of other Local~Group dSphs, dwarf
ellipticals, and dwarf irregulars.  In contrast to four out of nine~Milky~Way
dSphs (the four that lie beyond $D_{\rm MW}=100$~kpc), none of the six~known
M31 dSphs appears to have a dominant intermediate-age population.
\end{abstract}


\section{Introduction}

Recently two groups have reported the independent discovery of new Local
Group dwarf galaxies, potential companions of M31.  Armandroff et~al.\
(1998a) used a digital filtering technique to survey digitized POSS-II plates
covering a 1550 deg$^2$ area surrounding M31 and found several candidate
dwarf galaxies.  They demonstrated that their first candidate, And\,V, is a
dwarf spheroidal (dSph) galaxy sufficiently close to M31 to be its satellite.
An additional candidate, named And\,VI following van den Bergh's (1974)
convention, was resolved into stars and found not to contain H{\sc ii}
regions (Armandroff et~al.\ 1998b,c).  The second group, Karachentsev \&
Karachentseva (1998a,b) visually inspected film copies of POSS-II plates and
detected four~potential new dwarf companions of M31.  The two~brightest,
granulated detections were deemed to be the most promising candidates.  These
were initially named Pegasus\,Dw and Cassiopeia\,Dw after their parent
constellations (Karachentsev \& Karachentseva 1998a,b), and later renamed
Peg\,dSph and Cas\,dSph by Tikhonov \& Karachentsev 1999).
Note, Peg\,dSph is equivalent to And\,VI and was discovered independently by
the two~groups.  The galaxy Cas\,dSph is located in an area for which no
POSS-II plate scan was available to Armandroff et~al.\ (1998a).  We will
adopt the designations Peg\,dSph and Cas\,dSph throughout this {\em Letter}.

These three newly discovered candidate M31 dSphs, And\,V, Peg\,dSph, and
Cas\,Dw, double the census of known M31 dSphs.  It is important to test
whether the M31 dSphs follow the same relations between absolute magnitude,
surface brightness, metallicity, $M/L$ ratio, and galactocentric distance as
the Milky Way dSph satellites?  Do environment and distance from M31 affect
their formation histories as one might expect from van den Bergh's (1994) ram
pressure/tidal stripping scenario?

This {\em Letter} presents results from Keck imaging of Peg\,dSph and
Cas\,dSph that address some of these questions; a preliminary version of the
results were presented by Grebel (1999).  Our Keck results are also compared
to the results from three~other recent/ongoing studies of a similar nature
targeting these galaxies (Tikhonov \& Karachentsev 1999; Hopp et~al.\ 1999;
Armandroff et~al.\ 1998c, 1999).

\section{Observations and Data Reduction}

In preparation for a stellar spectroscopy program of M31's dSph companions,
Peg\,dSph and Cas\,dSph were imaged with the Keck~II 10-m telescope in
August--September~1998, using the Low Resolution Imaging Spectrograph (Oke
et~al.\ 1995) with a $2048^2$ CCD and a scale of $0\farcs215$~pixel$^{-1}$.
Short exposures were obtained in the $V$ and $I$ bands: $1\times300\,$s and
$3\times200\,$s, respectively, per galaxy.  Observing conditions were
photometric and the seeing ranged from 0\farcs7--1\farcs0 (FWHM).  The
overlapping field of view of the final, combined images is
$4\farcm6\times7\farcm1$.  Resolved objects (galaxies) and regions around
bright, saturated stars are excluded from further analysis (cf.~Reitzel
et~al.\ 1998).  Stellar photometry is carried out with the DAOPHOT/ALLFRAME
programs (Stetson 1994).  The photometric transformation uses airmass
coefficients from Krisciunas et~al.\ (1987) and observations of faint
secondary standards (Stetson, priv.\ comm.) in Landolt (1992) fields.

\section{Morphological Appearance}

Peg\,dSph and Cas\,dSph have the appearance of typical low-surface brightness 
spheroidal galaxies without recognizable concentrations of bright stars,
H\,{\sc ii} regions, or obvious globular clusters (Figure~1).  Peg\,dSph is not
detected in H$\alpha$, indicating the absence of currently star-forming
regions (Armandroff 1998a).  An ellipticity of~0.3 is estimated for both
galaxies, along with position angles of $160^{\circ}\pm20^{\circ}$ and
$55^{\circ}\pm20^{\circ}$ for Peg\,dSph and Cas\,dSph, respectively.  Hopp
et~al.'s (1999) measurement of the Peg\,dSph structural parameters is in
agreement with this estimate.  The ellipticities of Peg\,dSph and Cas\,dSph are
similar to those of Galactic dSph satellites at larger distances from the
Milky Way (e.g.,~Caldwell et~al.\ 1992), suggesting little or no distortion
through tidal effects.  A detailed analysis of the galaxies' structural
parameters will be presented in a later paper (Guhathakurta \& Grebel 1999).

\section{Stellar Luminosity Functions and Distances}

The cumulative stellar $I$-band luminosity function (LF) for both galaxies is
shown in Figure~2.  The LF of Cas\,dSph rises at $I\approx20.7\pm0.1$, and
this is interpreted as the tip of the red giant branch (TRGB).  The LF upturn
is more gradual for Peg\,dSph with $I_{\rm TRGB}\approx20.7\pm0.1$.  The
Galactic line-of-sight reddening is estimated from DIRBE/IRAS maps (Schlegel
et~al.\ 1998) to be $E_{B-V}=0.04\pm0.01$ and $0.17\pm0.03$, or
$A_I=0.09\pm0.01$ and $0.35\pm0.06$, for Peg\,dSph and Cas\,dSph, respectively.
An absolute magnitude of $M_I=-4$ for the TRGB (Lee at al.\ 1993) implies
true distance moduli of $(m-M)_0=24.6\pm0.2$ and $24.4\pm0.2$ for Peg\,dSph
and Cas\,dSph, respectively, where the errors include uncertainties in the
exact location of the TRGB (Poisson error, possible presence of
intermediate-age stars brighter than the TRGB), reddening, and
photometric calibration.

Tikhonov \& Karachentsev (1999) find apparent distance moduli for Peg\,dSph and
Cas\,dSph that translate into true distance moduli of $(m-M)_0=24.52$ and~24.36,
respectively, when the above foreground extinction correction is applied.
Corrected in the same manner, Hopp et~al.'s (1999) finding corresponds to
$(m-M)_0=24.43\pm0.2$ for Peg\,dSph.  These results are in excellent agreement
with the results of this study, when the uncertainties in the measurements
are taken into account.

\section{Color-Magnitude Diagrams and Metallicity Estimates}

The color-magnitude diagrams (CMDs) of Peg\,dSph and Cas\,dSph show prominent red
giant branches (RGBs) and no evidence for young blue stars (Figure~3).  Note,
the surrounding $R>3'$ ``field'' CMDs in Figure~3, each drawn from a region
whose area is about 45\% that of the $R<3'$ galaxy area.  The field CMDs are
slightly contaminated by stars in the outskirts of the dSph galaxies ($<10\%$
of the number of stars in the corresponding $R<3'$ galaxy CMD), but this is
unimportant for the purposes of this study since no significant radial
gradient appears to be present in the galaxies' properties.  The Cas\,dSph RGB
is more densely populated than that of Peg\,dSph, a reflection of Cas\,dSph's
higher stellar surface density.  A small metal-rich
($\rm[Fe/H]<-1$) population appears to be indicated by stars with $V-I\ge2$
near the TRGB in Cas\,dSph (Figure~3).  The density and color distribution of
stars above the Cas\,dSph TRGB are similar to that in the corresponding
``field'' CMD, but the presence of a small intermediate-age asymptotic giant
branch (AGB) population cannot be ruled out.  The Peg\,dSph CMD displays an
excess of stars above the TRGB relative to the field CMD; this is also seen
as a tail in the LF brighter than the TRGB.  These super-TRGB stars may
indicate the presence of a significant intermediate-age AGB population in
Peg\,dSph; they are not likely to be blend artifacts (cf.~Grillmair et~al.\
1996; Mart\'{\i}nez-Delgado \& Aparicio 1997).

The mean metallicity is computed from the median reddening-corrected color,
$(V-I)_0$, at $M_I=-3.5$ (Armandroff et~al.\ 1993; Caldwell et~al.\ 1998),
computed over the range $1<(V-I)_0<2$.  Outside this color range the surface
density of objects is identical for the inner $R<3'$ and surrounding $R>3'$
regions, indicating that the majority of these objects are foreground or
background sources.  For Peg\,dSph, $(V-I)_{0,-3.5}=1.46\pm0.02$ for
$21.1\leq{I}\leq21.3$, which yields $\rm\langle[Fe/H]\rangle=-1.35\pm0.07$.
For Cas\,dSph, $(V-I)_{0,-3.5}=1.43\pm0.01$ for $21.1\leq{I}\leq21.3$, which
yields $\rm\langle[Fe/H]\rangle=-1.42\pm0.04$.  These errors are the formal
errors of the mean; the overall uncertainty in the mean metallicities is
conservatively estimated to be $\pm0.3$~dex due to systematic errors in
photometry/calibration.  These metallicity estimates are in good agreement
with those derived from the standard globular cluster RGB fiducials (Da Costa
\& Armandroff 1990) shown in Figure~3.

The observed width of the RGB in the Peg\,dSph and Cas\,dSph CMDs,
$\delta(V-I)\approx0.4$, corresponds to a metallicity spread of roughly
1~dex.  The formal uncertainty in the color measurement (see Fig.~3) is
significantly smaller than the observed spread in bright RGB color.  The
effect of systematic errors (e.g.,~crowding, flat fielding), however, may be
as large as $\sigma(V-I)\sim0.1$, judging from a comparative analysis of a
subset of the data obtained in poorer seeing and at a different CCD
orientation with respect to the primary data set.  The fact that the RGB
widths are similar for Peg\,dSph and Cas\,dSph despite the substantial
difference in their stellar surface density argues against the widths being
caused by crowding-related photometric error.  Finally, the apparent flaring
of the Cas\,dSph RGB near its bright tip is strongly suggestive of a spread
in [Fe/H].

Measurements of the mean metallicity in three~radial bins ($R<1'$, $1'<R<2'$,
and $2'<R<3'$), based on a wider $I$-magnitude range ($\Delta{I}=0.5$~mag),
places a robust upper limit on any radial metallicity gradient in the
galaxies.  The variation in median $(V-I)_0$ color over the inner $3'$ region
is found to be $<0.03$~mag corresponding to a variation in mean metallicity
of $<0.05$ dex, comparable to the formal error in $\rm\langle[Fe/H]\rangle$
quoted above.  Thus Peg\,dSph and Cas\,dSph show no evidence for a radial
metallicity gradient, in contrast with what is seen in some dSphs with
substantial intermediate-age populations (see Grebel 1997 for a summary).

\section{Central Surface Brightness}

The central surface brightness in the $V$ band, measured in a $40''$-diameter
aperture after subtraction of bright foreground stars and corrected for
line-of-sight extinction (same procedure as in Armandroff et~al.\ 1998a), is
$\mu_{V,0}\approx24.5$ and 23.6~mag~arcsec$^{-2}$ for Peg\,dSph and Cas\,dSph,
respectively; these brightnesses are higher than the observed $\mu_{V,0}$
values for the other M31 dSphs (Caldwell et~al.\ 1992; Armandroff et~al.\
1998a): 24.9 (And\,I), 24.8 (And\,II), 25.3 (And\,III), and
25.2~mag~arcsec$^{-2}$ (And\,V).  Peg\,dSph and Cas\,dSph, along with the
newly discovered And\,V dSph (Armandroff et~al.\ 1998a), follow the general
$\rm\langle[Fe/H]\rangle$-vs-$\mu_{V,0}$ relationship defined by Local Group
dwarf ellipticals (dEs), dwarf irregulars, and dSphs (Figure~4, {\em upper
panel\/}).  Absolute $B$-band magnitudes, $M_B$, are computed from the total
apparent magnitudes, $B_T$, estimated by Karachentsev \& Karachentseva
(1998a).  Peg\,dSph lies on the  mean $M_B$-vs-$\mu_{V,0}$ relation defined
by other Local Group dwarf galaxies (Figure~4, {\em lower panel\/}), while
Cas\,dSph is somewhat less luminous than other dwarfs of comparable
$\mu_{V,0}$.

\section{Discussion}

A summary of the properties of the two new dSphs Peg\,dSph and Cas\,dSph derived
from Keck imaging is given in Table~1.  Their mean metallicities resemble
that of the M31 globular cluster system ($\rm[Fe/H]=-1.2$---Huchra et~al.\
1991) and of the intermediate-age/old field populations in M31's dE
companions, but are higher than the $\rm\langle[Fe/H]\rangle$ of the other
four~M31 dSphs (cf.~Grebel 1997; Armandroff et~al.\ 1998a).  Peg\,dSph and
Cas\,dSph also have the highest central surface brightnesses of the known M31
dSphs.  Their detection may have been hampered by the presence of nearby,
bright foreground stars.

Heliocentric distances of $830\pm80$~kpc for Peg\,dSph and $760\pm70$ kpc for
Cas\,dSph translate into distances of $280\pm85$~kpc and $215\pm75$~kpc from
M31, respectively, adopting the Cepheid-based distance of $770\pm30$~kpc to
M31 (cf.~Freedman \& Madore 1990; Holland 1998).  Thus Peg\,dSph and
Cas\,dSph are the most distant of the currently known M31 dSph satellites.
The Leo\,I and Leo\,II
dSphs have comparable distances from the Milky Way, lending credence to
Karachentsev \& Karachentseva's (1998a) suggestion that Peg\,dSph and Cas\,dSph
belong to an extended subsystem of M31 companions together with And\,V and
LGS\,3.

The fraction of stars belonging to an intermediate-age population shows a
bimodal distribution among the Milky Way dSphs, with small intermediate-age
fractions at small Galactocentric distances and dominant fractions at large
distances (van den Bergh 1994).  By contrast, all the M31 dSph companions
appear to have small intermediate-age populations (with the possible
exception of And\,II---Da Costa 1998), even though they are located at
comparable distances from their more massive parent galaxy.  The possibility
of highly eccentric orbits in the M31/M33 system cannot be ruled out.  Our
planned spectroscopic observations will help constrain the orbits of the M31
dSphs through radial velocity measurements and should lead to more accurate
metallicity and abundance spread determinations.

\acknowledgements

We thank Igor Karachentsev, Valentina Karachentseva, Taft Armandroff, and
George Jacoby for helpful discussions, and an anonymous referee for useful
comments.  Mike Bolte kindly provided IRAF and DAOPHOT scripts for reduction
of the Keck data.  We are indebted to Peter Stetson for making available his
secondary standards, without which a reliable calibration would not have been
possible.  EKG gratefully acknowledges support by Dennis Zaritsky through
NASA LTSA grant NAG-5-3501 and by NASA through grant HF-01108.01-98A from the
Space Telescope Science Institute, which is operated by the Association of
Universities for Research in Astronomy, Inc., under NASA contract NAS5-26555.

\newpage

%
%
%
%
\figcaption{Negative grayscale representation of $V$-band images of
Pegasus\,dSph and Cassiopeia\,dSph ($1\times300$\,s for each galaxy) obtained
with the Low Resolution Imaging
Spectrograph on the 10-m Keck~II telescope in August~1998.  The seeing FWHM
for these observations was 0\farcs7--1\farcs0.
\label{fig1}}
 
%
%
\figcaption{Cumulative $I$-band luminosity functions (LFs) of the central
$R<3'$ regions of Peg\,dSph and Cas\,dSph and of the surrounding $R>3'$ field
regions (scaled to match the area of the inner $R<3'$ region).  The arrows 
mark the adopted position of the tip of the red giant branch (TRGB) in the
field-star subtracted LF (dotted line).  The bright end of the field-star
subtracted LF appears to extend to slightly brighter magnitudes than the
adopted TRGB location, especially for Peg\,dSph; this may indicate the presence
of luminous AGB stars. 
\label{fig2}}

%
%
\figcaption{Color-magnitude diagrams (CMDs) of the inner $R<3'$ region of
Peg\,dSph and Cas\,dSph, with their corresponding surrounding $R>3'$ field
regions
(each field region has roughly 45\% the area of the inner $R<3'$ region).
The error bands in the right panels represent the formal (DAOPHOT)
$\pm1\sigma$ error in $V-I$ color; the overall errors are likely to be
somewhat larger.
Note, the ``field'' CMD is slightly contaminated by red giant stars in the
outer regions ($>60$~pc) of the dSphs.  The solid grey lines in the left
panels are globular cluster
fiducials from Da Costa \& Armandroff (1990), appropriately reddened and
shifted to the distance moduli of the galaxies, for a range of $\rm[Fe/H]$
values: (left to right) M15 ($-2.2$), NGC\,6752 ($-1.5$), NGC\,1851 ($-1.2$),
and 47\,Tuc ($-0.7$).
\label{fig3}}

%
%
\figcaption{({\bf a})~Mean metallicity, $\rm\langle[Fe/H]\rangle$, plotted
versus central surface brightness in the $V$ band, $\mu_{V,0}$, for nearby
dwarf galaxies.  The data are from Grebel (1997), Armandroff et~al.\ (1998a),
and the present study.  Peg\,dSph and Cas\,dSph (two~circled dots) follow the
mean $\rm\langle[Fe/H]\rangle$-vs-$\mu_{V,0}$  relation within the
errors.~~~
({\bf b})~Absolute $B$-band magnitude, $M_B$, versus central surface
brightness for the same galaxies that are plotted in~(a) with the exception
of the And\,V dSph for which an $M_B$ measurement is unavailable.  The $M_B$
values are uncertain at the level of $\pm0.5$~mag (Karachentsev, priv.\
comm.).  The data are from Mateo (1998) and the present study.
\label{fig4}}

\newpage
 
%
%

\begin{deluxetable}{lcc}
\tablecaption{Properties of Peg\,dSph and Cas\,dSph\label{table_1}}
\footnotesize
\tablehead{\colhead{Parameter} & \colhead{Peg\,dSph}    & \colhead{Cas\,dSph}}
\startdata
$\alpha$ (J2000)  & $\rm23^h51^m39^s$      &  $\rm23^h26^m31^s$     \\
$\delta$ (J2000)  & $24^{\circ}35'42''$ &  $50^{\circ}41'31''$\\
{\em l, b}        & $106.0^{\circ}$, $-36.3^{\circ}$ &
                    $109.5^{\circ}$,  $-9.9^{\circ}$ \\
Position angle    & $160^{\circ}\pm20^{\circ}$ & $55^{\circ}\pm20^{\circ}$\\
Ellipticity       & $0.3\pm0.1$        &  $0.3\pm0.1$       \\
$E_{B-V}$       & $0.04\pm0.01$   &  $0.17\pm0.03$ \\
$(m-M)_0$         & $24.6\pm0.2$   &  $24.4\pm0.2$ \\
Distance$_{\rm MW}$ & $830\pm80$~kpc    &  $760\pm70$~kpc   \\
Distance$_{\rm M31}$& $280\pm85$~kpc & $215\pm75$~kpc\\
Distance$_{\rm M33}$&$340\pm90$~kpc & $ 430\pm80$~kpc\\
$B_T$         & $14.5\pm0.5^{\rm a}$  &  $16.0\pm0.5^{\rm a}$\\
$V_T$         & $14.1\pm0.2^{\rm b}$  &  \\
$\mu_{V,0}$       & $24.5\pm0.2$~mag~arcsec$^{-2}$ &
$23.6\pm0.2$~mag~arcsec$^{-2}$\\
$\rm\langle[Fe/H]\rangle$        & $-1.3\pm0.3$    & $-1.4\pm0.3$\\
\tablenotetext{a}{From Karachentsev \& Karachentseva (1998).}
\tablenotetext{b}{From Hopp et~al.\ (1999), corrected using Schlegel et~al.'s
(1998) extinction estimate.}
\enddata
\end{deluxetable}


\begin{thebibliography}{}

\bibitem{}
Armandroff, T.\,E., Da Costa, G.\,S., Caldwell, N., \& Seitzer, P. 1993, AJ, 
106, 986
\bibitem{}
Armandroff, T.\,E., Davies, J.\,E., \& Jacoby, G.\,H. 1998a, AJ, 116, 2287 
\bibitem{}
Armandroff, T.\,E., Davies, J.\,E., \& Jacoby, G.\,H. 1998b, in Dwarf
Tales, eds.\ E.\ Brinks \& E.\,K.\ Grebel, Vol.\ 3, p.\ 2 
\bibitem{}
Armandroff, T.\,E., Davies, J.\,E., \& Jacoby, G.\,H. 1998c, in IAU Coll.\
171, The Low Surface Brightness Universe, eds.\ J.\,I.\ Davies, C.\ Impey, \&
S.\ Philipps (San Francisco: ASP), in press 
\bibitem{}
Armandroff, T.\,E., Davies, J.\,E., \& Jacoby, G.\,H. 1999, in preparation
\bibitem{}
Caldwell, N., Armandroff, T.\,E., Seitzer, P., \& Da Costa, G.\,S. 1992, AJ,
103, 840
\bibitem{}
Caldwell, N., Armandroff, T.\,E., Da Costa, G.\,S., \& Seitzer, P. 1998, AJ, 
115, 535
\bibitem{}
Da Costa, G.\,S., \& Armandroff, T.\,E. 1990, AJ, 100, 162
\bibitem{}
Da Costa, G.\,S. 1998, in Stellar Astrophysics for the Local Group: A
First Step to the Universe, eds.\ A.\ Aparicio \& A.\ Herrero
(Cambridge: Cambridge University Press), in press
\bibitem{}
Freedman, W.\,L., \& Madore, B.\,F. 1990, ApJ, 365, 186
\bibitem{}
Grebel, E.\,K. 1997, Reviews in Mod.\ Astron., 10, 29
\bibitem{}
Grebel, E.\,K. 1999, in IAU Symp.\ 192, The Stellar Content of the Local
Group, eds.\ P.\ Whitelock \& R.\ Cannon (San Francisco: ASP), in press
\bibitem{}
Grillmair, C.\,J., Lauer, T.\,R., Worthey, G., et~al. 1996, AJ, 112, 1975
\bibitem{}
Guhathakurta, P., \& Grebel, E.\,K. 1999, ApJ, in preparation
\bibitem{}
Holland, S. 1998, AJ, 115, 1916
\bibitem{}
Hopp, U., Schulte-Ladbeck, R.\,E., Greggio, L., \& Mehlert, D. 1999, 
A\&AL, submitted
\bibitem{}
Huchra, J.\,P., Kent, S.\,M., \& Brodie, J.\,P. 1991, ApJ, 370, 495
\bibitem{}
Karachentsev, I.\,D., \& Karachentseva, V.\,E. 1998a, in Dwarf Tales,
eds.\ E.\ Brinks \& E.\,K.\ Grebel, Vol.\ 3, p.\ 1 
\bibitem{}
Karachentsev, I.\,D., \& Karachentseva, V.\,E. 1998b, A\&A, submitted
\bibitem{}
Krisciunas, K., Sinton, W., Tholen, K., et~al. 1987, PASP, 99, 887
\bibitem{}
Landolt, A.\,U. 1992, AJ, 104, 340
\bibitem{}
Lee, M.\,G., Freedman, W.\,L., \& Madore, B.\,F. 1993, ApJ, 417, 553
\bibitem{}
Mateo, M. 1998, ARA\&A, 36, 435
\bibitem{}
Mart\'{\i}nez-Delgado, D., \& Aparicio, A. 1997, ApJ, 480, 107
\bibitem{}
Oke, J.\,B., Cohen, J.\,G., Carr, M., et~al. 1995, PASP, 107, 375
\bibitem{}
Reitzel, D.\,B., Guhathakurta, P., \& Gould, A. 1998, AJ, 116, 707
\bibitem{}
Schlegel, D.\,J., Finkbeiner, D.\,P., \& Davis, M. 1998, ApJ, 500, 525
\bibitem{}
Stetson, P.\,B. 1994, PASP, 106, 250
\bibitem{}
Tikhonov, N.\,A., \& Karachentsev, I.\,D. 1999, AstL, submitted
\bibitem{}
van den Bergh, S. 1974, ApJ, 191, 271 
\bibitem{}
van den Bergh, S. 1994, AJ, 108, 2145

\end{thebibliography}
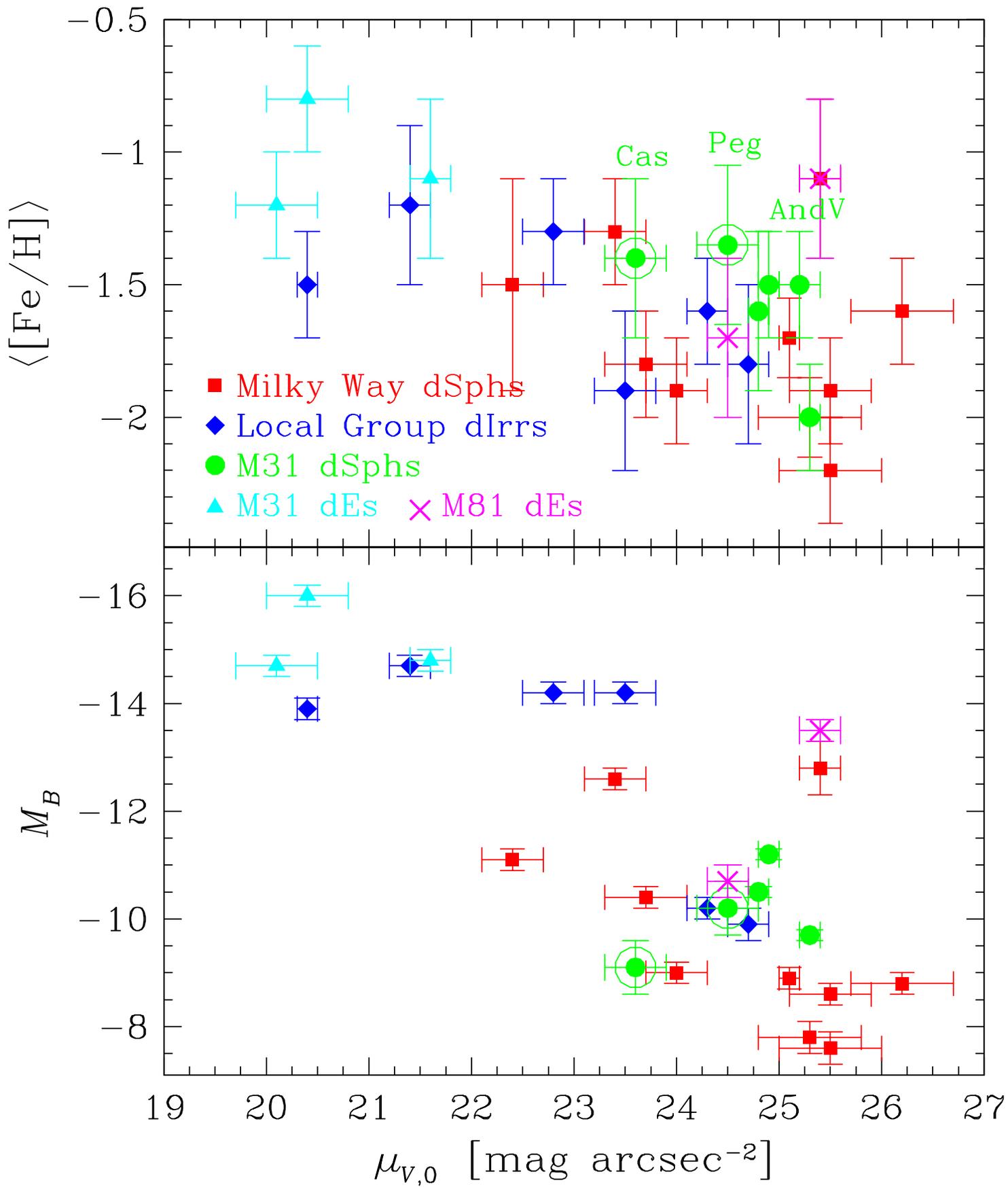
\end{document}